\documentclass[aps,prb,twocolumn,amsmath,amssymb,superscriptaddress,floatfix,eqsecnum]{revtex4-2}

\usepackage{float}
\usepackage{physics}
\usepackage{times}
\usepackage{bm}
\usepackage{textcomp}
\usepackage{graphicx} 
\usepackage{braket}
\usepackage{array}
\usepackage[breaklinks]{hyperref}
\usepackage{subdepth}
\usepackage{mathtools}
\usepackage{cancel,soul,ulem}
\usepackage{wasysym}

\hypersetup{colorlinks=true, linkcolor=blue, citecolor=blue, filecolor=blue, urlcolor=blue}

\newcommand{\RN}[1]{\textup{\uppercase\expandafter{\romannumeral#1}}}

\begin{document}

\title{Single monkey-saddle singularity
of a Fermi surface and its instabilities}

\author{\"{O}mer M. Aksoy}
\affiliation{Condensed Matter Theory Group, Paul Scherrer Institute, CH-5232 Villigen PSI, Switzerland}

\author{Anirudh Chandrasekaran}
\affiliation{Department of Physics and Centre for the Science of Materials,
Loughborough University,  Loughborough LE11 3TU, UK}

\author{Apoorv Tiwari}
\affiliation{Department of Physics, KTH Royal Institute of Technology, 106 91  Stockholm, Sweden}

\author{Titus Neupert}
\affiliation{University of Zurich, Winterthurerstrasse 190, 8057 Zurich, Switzerland}

\author{Claudio Chamon}
\affiliation{Department of Physics, Boston University, Boston, Massachusetts, 02215, USA}

\author{Christopher Mudry}
\affiliation{Condensed Matter Theory Group, Paul Scherrer Institute, CH-5232 Villigen PSI, Switzerland}
\affiliation{Institut de Physique, EPF Lausanne, CH-1015 Lausanne, Switzerland}

\date{\rm\today}

\begin{abstract}
Fermi surfaces can undergo sharp transitions under smooth changes of
parameters. Such transitions can have a topological character, as is
the case when a higher-order singularity, one that requires cubic
or higher-order terms to describe the electronic dispersion near the
singularity, develops at the transition. When time-reversal and
inversion symmetries are present, odd singularities can only appear
in pairs within the Brillouin zone. In this case, the combination of
the enhanced density of states that accompanies these singularities
and the nesting between the pairs of singularities leads to
interaction-driven instabilities.  We present examples of single
$n=3$ (monkey-saddle) singularities when time-reversal and inversion
symmetries are broken. We then turn to the question of what
instabilities are possible when the singularities are isolated.  For
spinful electrons, we find that the inclusion of repulsive
interactions destroys any isolated monkey-saddle singularity present
in the noninteracting spectrum by developing Stoner or Lifshitz
instabilities.  In contrast, for spinless electrons and at the
mean-field level, we show that an isolated monkey-saddle singularity
can be stabilized in the presence of short-range repulsive
interactions.
\end{abstract}

\maketitle

%%%%%%%%%%%%%%%%%%%%%%%%%%%%%%%%%%%%%%%%%%%%%%%%%%%%%%%%%%%%%%%%%%%%%%%%%%

\section{Introduction}
\label{sec: Introduction}

Topological transitions of Fermi surfaces are currently a topic of
active
research~\cite{Liu10,Okamoto10,HuxleyLifshitz11,Khan14,Benhabib15,
Slizovskiy15,AokiCelrln16,Volovik2017,monkey_saddle,
Slizovskiy2018,Mohan2018,Sherkunov18,Barber19,yuan2019magic,
supermetal,Rao2020,HOS_RG_Chubkov,Lin2020,Oriekhov2021,Guerci2022,Seiler2022}.
This is particularly so when space is two dimensional, in which case
they are often associated with band singularities that cause the
density of states (DOS) to diverge. To be precise, in a Fermi-surface
topological transition~\cite{Lifshitz60}, the topology of the Fermi
surface undergoes a sudden change upon tuning some
parameters. At the transition, the Fermi surface may develop a
singularity due to the presence of one or more saddle points in the
dispersion. A saddle point is responsible for a divergent DOS, which
in turn may lead to many distinct physical phenomena such as
charge and spin order, superconductivity, and diverging
susceptibilities.

In two-dimensional space,
an ordinary saddle, known as the van Hove singularity,
can be subsumed as the quadratic dispersion
$\varepsilon(\bm{k})\propto k^{2}_{x} - k^{2}_{y}$ that causes a
logarithmic divergence of the DOS at the Fermi level
$\varepsilon^{\,}_{\mathrm{F}}=0$.
Higher-order singularities, in contrast,
are characterized by a $\bm{k}\cdot\bm{p}$  expansion in which the
lowest-order terms are higher than quadratic.
For example,
$\varepsilon(\bm{k})\propto k^{3}_{x}-3\,k^{\,}_{x}\,k^{2}_{y}$
implies a singular DOS at the Fermi level
$\varepsilon^{\,}_{\mathrm{F}}=0$ of order $n=3$.
These cause power-law divergences of the DOS. In the
context of band theory in two-dimensional space,
higher-order singularities have been
classified using sets of integer indices, based on
symmetry, scaling, number of relevant perturbations,
etc.\
\citep{classification_catastrophe,classification_MIT}. Furthermore,
their intimate connection with high-symmetry points in the Brillouin
zone (BZ) has also been worked
out\,
\citep{classification_catastrophe}. 

A divergent DOS leads to a subtle competition between enhanced
electron-electron interactions
on the one hand, and enhanced screening of
interactions on the other hand\,
\citep{DOS_screening,gonzalez1997renormalization,Gonzalez03,evHs1}.
Combined with the non-trivial band
geometry, higher-order singularities may activate one or more
instability channels, especially when they are nested or when they
occur in symmetry-related positions in the BZ. The presence of a
single higher-order singularity at the Fermi level is also
expected to lead to a breakdown of Fermi-liquid theory, in the
presence of interactions\,
\citep{monkey_saddle,supermetal}.
A number of other recent works also seem to indicate marginal
Fermi-liquid behavior for systems with even higher-order
singularities\,
\citep{Ruthenate_MFL,MFL_in_TBG}. For example, the
$T$-linear dependence of resistivity in twisted bilayer graphene
has been explained as a consequence of the marginal Fermi-liquid
behavior arising from the electrons in the vicinity of an extended
van Hove singularity\,
\citep{Ruthenate_MFL}. Marginal Fermi-liquid
behavior has also been associated with $\text{Sr}^{\,}_{3} \text{Ru}^{\,}_{2}
\text{O}_7$\,
\citep{Ruthenate_MFL}, and proposed to arise from
two-electron scattering processes in which electrons from a cold
region (non-singular region) scatter into a pair of states, one in the
cold region and another in the hot region (i.e., region near a
higher-order singularity). It is 
worth mentioning that
$\text{Sr}^{\,}_{3} \text{Ru}^{\,}_{2} \text{O}_7$
hosts a $n=4$
rotationally symmetric saddle,
\footnote{
In the earlier literature, higher-order singularities, with
their power-law diverging DOS, were recognized as objects distinct from
the conventional van Hove singularity. The somewhat extended (and
asymmetric) nature of the contours of the higher-order saddle
$\varepsilon(\bm{k})= k^{4}_{x} - k^{2}_{y}$ appears to have
motivated the name ``extended van Hove singularity'' (see
Refs.\ \cite{evHs1,evHs2,evHs3,evHs4,evHs5,evHs6,evHs7}).
         } the latter having been analyzed in Refs.\
\cite{ruthenate_HOS_Chamon_Shtyk,monkey_saddle,supermetal}.

To reach non-Fermi-liquid behavior in such systems, it is imperative
to try to avoid instabilities towards symmetry-broken phases. In this
regard, when singularities appear in pairs, at symmetry-related points
in the Brillouin zone, scattering between states at the two points
will generically stabilize a symmetry-broken phase at low
temperatures\,
\citep{monkey_saddle}. Even singularities may appear
alone at a high-symmetry point that maps onto itself under
time-reversal symmetry, but an odd higher-order singularity cannot.

In this paper, we present two single-particle Hamiltonians in Sec.\
\ref{sec:models} that encode the kinetic energy of non-interacting
electrons constrained to move in two-dimensional space.  By explicitly
breaking time-reversal and inversion symmetries so as to avoid the
doubling of the number of higher-order singular points that occur
when these symmetries hold, we obtain a single Fermi ``surface'' with
a single odd higher-order singularity.  More precisely, by tuning one
parameter, both models are made to host the three-fold-rotationally
symmetric saddle of order $n=3$,
also known as the monkey saddle.
One of the two models is a deformation of Haldane's Chern insulator
on the honeycomb
lattice\, \cite{Haldane_model} through the addition of a staggered
chemical potential (see Ref.\ \cite{Chandrasekaran2023}). 
By tuning the staggered chemical potential, a
monkey-saddle singularity appears at just one of the two inequivalent
corners of the Brillouin zone.  Furthermore, it is possible to tune
the ratio of the next-nearest- to nearest-neighbor hoppings so that
the energy of the monkey saddle is smaller in absolute value than that
at the non-equivalent corner of the Brillouin zone. In this regime,
the anomalous Hall conductivity is nonvanishing, but it contains no
singular behavior other than that coming from the DOS.  We then turn
our attention to the role played by interactions in Secs.~\ref{sec:
  The monkey saddle realizes a non-Fermi liquid} and \ref{sec: Mean
  field analysis}.  For spinful electrons, when the Fermi energy
matches that of the monkey saddle in the noninteracting limit, we show
that the presence of short-range repulsive interactions always leads
to the disappearance of an isolated monkey-saddle singularity within a
mean-field approximation.  For spinless electrons, we show that a
monkey-saddle singularity can be stabilized in the presence of
repulsive interactions at the mean-field level, but with renormalized
parameters (compared to those for which the singularity appears in the
absence of interactions).  We summarize the results in Sec.\ \ref{sec:
  Summary}.

\section{Models}
\label{sec:models}

In this section, we construct two single-particle dispersions
each of which hosts a single higher-order singularity of odd parity,
namely, the monkey saddle defined by
\begin{equation}
\varepsilon_{\mathrm{ms}}(\bm{k}):=
\alpha
\left(
k^{3}_{x}-3k^{\,}_{x}\,k^{2}_{y}
\right)=
\alpha\,k^{3}\, \cos(3\theta),
\label{eq: monkey saddle dispersion}
\end{equation}
where the last equality corresponds to writing the dispersion in polar
coordinates with respect to the singular point.  The constant
$\alpha$ has units of energy times length cubed.

\subsection{Topological insulator surface state}

We modify a previously derived
$\bm{k}\cdot\bm{p}$ model for the surface states of
$\text{Bi}^{\,}_{2} \text{Te}^{\,}_{3}$\,
\citep{TI_surface_state_Bi2Te3}
by adding a term to the Hamiltonian that explicitly breaks
time-reversal symmetry.
This allows us to obtain a single monkey
saddle at the $\Gamma$ point under appropriate tuning.

First, we briefly review the original model
for the surface states of
$\text{Bi}^{\,}_{2} \text{Te}^{\,}_{3}$.
A minimal $\bm{k}\cdot\bm{p}$ theory can be
constructed for the system by symmetry arguments. Total angular
momentum $1/2$ is manifest as a spinor degree of freedom, giving rise
to two bands. The symmetries in the system form a group obtained
by taking the semi-direct product of
the cyclic group generated by the $2\pi/3$ rotation,
the cyclic group generated by the
reflection $x\to-x$,
and the cyclic group generated by reversal of time $t\to-t$.
When acting on the
``spin'' degree of freedom these symmetries are represented
by
\begin{subequations}
\begin{equation}
\mathcal{R}\equiv e^{+\mathrm{i} \frac{\pi}{3} \sigma^{\,}_{z}},
\qquad
\mathcal{M}\equiv \mathrm{i} \sigma^{\,}_{x},
\qquad
\mathcal{T}\equiv \mathrm{i} \sigma^{\,}_{y} \mathsf{K},
\end{equation}
respectively,
where
$\mathsf{K}$ denotes complex conjugation and we introduced the
three Pauli matrices
$\bm{\sigma}=(\sigma_{x},\sigma_{y},\sigma_{z})$ acting
on the spinor components. Their combined actions on two-dimensional momentum space
that we parametrize with the coordinates
$k^{\,}_{\pm} = k^{\,}_{x} \pm \mathrm{i} k^{\,}_{y}$
with the $\Gamma$ point as the origin
and
``spin'' space parametrized with the coordinates
$\sigma^{\,}_{\pm}=\sigma^{\,}_{x}\pm\mathrm{i}\sigma^{\,}_{y}$
and
$\sigma^{\,}_{z}$,
are
\begin{align}
&
\mathcal{R}:
\left\{
\begin{array}{l}
k_{\pm} \mapsto e^{\pm\mathrm{i} 2\pi /3}\,k^{\,}_{\pm},
\\%\\
\sigma_{\pm}\mapsto e^{\pm\mathrm{i} 2\pi /3}\,\sigma^{\,}_{\pm},
\qquad
\sigma^{\,}_{z}\mapsto \sigma^{\,}_{z},
\end{array}
\right.
\\
&
\mathcal{M}:
\left\{
\begin{array}{l}
k^{\,}_{\pm}\mapsto-k^{\,}_{\mp},
\\%\\
\sigma^{\,}_{\pm}\mapsto\sigma^{\,}_{\mp},
\qquad
\sigma^{\,}_{z}\mapsto-\sigma^{\,}_{z},
\end{array}
\right.
\\
&
\mathcal{T}:
\left\{
\begin{array}{l}
k^{\,}_{\pm}\mapsto-k^{\,}_{\mp},
\\%\\
\sigma^{\,}_{\pm}\mapsto-\sigma^{\,}_{\mp},
\qquad
\sigma^{\,}_{z}\mapsto-\sigma^{\,}_{z},
\end{array}
\right.
\end{align}
\end{subequations}
respectively. 
The dependence on momentum $\bm{k}\in\mathbb{R}^{2}$
of the most general single-particle
two-band Hamiltonian that is symmetric under
$\mathcal{R}$,
$\mathcal{M}$,
and $\mathcal{T}$
is then given by
\begin{equation}
\begin{split}
H^{\,}_{\mathrm{sym}}(\bm{k}) \equiv &\,
\left(
-\mu
+
\frac{k^{2}}{2m^{*}}
+
c^{\,}_{1}\, k^{4}
\right)
\openone
\\
&
+
\frac{\mathrm{i}v}{2}
\left(
1
+
c^{\,}_{2}\,k^{2}
\right)
\left(
k^{\,}_{+}\,\sigma^{\,}_{-}
-
k^{\,}_{-}\,\sigma^{\,}_{+}
\right)
\\
&
+
\frac{c^{\,}_{3}}{2}
\left(
k^{3}_{+}
+
k^{3}_{-}
\right)
\sigma^{\,}_{z},
\end{split}
\end{equation}
up to quartic order in an expansion of the momenta measured relative
to the $\Gamma$ point. This single-particle Hamiltonian depends on
the six real-valued dimensionful couplings
$\mu$,
$m^{*}$,
$c^{\,}_{1}$,
$v$,
$c^{\,}_{2}$,
and
$c^{\,}_{3}$.
Adding a Zeeman term, whose strength is parametrized by
the real-valued dimensionful coupling $b$,
and using polar coordinates delivers
\begin{equation}
\begin{split}
H(\bm{k}) \equiv &\,
\left(
-\mu
+
\frac{k^{2}}{2m^{*}}
+
c^{\,}_{1}\,k^{4}
\right)
\openone
\\
&
+
v
\left(
1
+
c^{\,}_{2}\,
k^{2}
\right)
k
\left[
\cos(\theta)\, \sigma^{\,}_{y}
-
\sin(\theta)\, \sigma^{\,}_{x}
\right]
\\
&
+
\left[
c^{\,}_{3}\,k^{3}\,\cos (3\theta)
+
b
\right]
\sigma^{\,}_{z}.
\end{split}
\label{eq:def first Hamiltonian}
\end{equation}
Hamiltonian (\ref{eq:def first Hamiltonian}) has the
single-particle dispersion
\begin{equation}
\begin{split}
\xi^{\,}_{\pm}(\bm{k})=&\,
-
\mu
+
\frac{k^{2}}{2m^{*}} + c^{\,}_{1}\, k^{4}
\\
&\,
\pm
\sqrt{
v^{2} \left(1 + c^{\,}_{2}\, k^{2} \right)^{2} k^{2}
+
\left[ c^{\,}_{3}\, k^{3} \cos (3\theta) + b \right]^{2}
     }.
\end{split}
\label{eq:def first Hamiltonian two dispersions}
\end{equation}

We expand this pair of dispersions up to
quartic order in the momenta
\begin{equation}
\begin{split}
\xi^{\,}_{{\mp}}(\bm{k}) \approx&\,
-\mu
{\mp}|b|
+
\frac{k^{2}}{2m^{*}}
\left(1 {\mp} \frac{m^{*} v^{2}}{|b|} \right)
\\
&\,
{\mp}
\mathrm{sgn}(b) \, c^{\,}_{3} \, k^{3} \cos \left( 3 \theta \right)
\\
&\,
{\pm}
\frac{v^{4} {\pm} 8 c^{\,}_{1}\, |b|^{3}-8 c^{2}_{2}\, v^{2}\, b^{2}}{8|b|^{3}}
\,
k^{4},
\end{split}
\label{eq:def first Hamiltonian lower dispersions quartic order}
\end{equation}
As promised, the monkey saddle appears in the ``$-$'' band upon tuning
the magnitude $|b|$ of the Zeeman term to the value $m^{*}\,v^{2}$,
thereby removing the $k^2$ term from Eq.~%
\eqref{eq:def first Hamiltonian lower dispersions quartic order}.
Henceforth, we work
in the two-dimensional region of parameter space for which
\begin{equation}
\xi^{\,}_{-}(\bm{k})=
-\mu
+ \alpha \,k^{3} \cos(3\theta)
+\mathcal{O}(k^{4})
\end{equation}
for $\alpha, \mu \in \mathbb{R}$.

\subsection{Haldane model}

We start from the single-particle tight-binding Hamiltonian
on the honeycomb lattice introduced by Haldane
in Ref.\ \onlinecite{Haldane_model}.
This single-particle tight-binding Hamiltonian
realizes a Chern insulator by breaking the time-reversal symmetry
and the three mirror symmetries of the point group $C^{\,}_{3v}$ of
the underlying triangular Bravais lattice.
We are going to show that
it also hosts a single monkey saddle at a Fermi level that
lies in the ``low-energy'' spectrum of the Hamiltonian.

We denote with A and B the two interpenetrating triangular sublattices
to the honeycomb lattice. Let
\begin{equation}
\bm{a}^{\,}_{1} =
\begin{pmatrix}1\\ \\0\end{pmatrix},
\quad
\bm{a}^{\,}_{2} =
\begin{pmatrix}-\frac{1}{2}\\ \\+\frac{\sqrt{3}}{2}\end{pmatrix},
\quad
\bm{a}^{\,}_{3} =
\begin{pmatrix}-\frac{1}{2}\\ \\-\frac{\sqrt{3}}{2}\end{pmatrix},
\label{eq:defs a1 a2 a3}
\end{equation}
denote the vectors that connect any site in sublattice A to its three nearest 
neighbors in sublattice B, where we have set the lattice spacing of the honeycomb lattice to unity.

Three of the six next-nearest-neighbor vectors
in the triangular sublattice A are given by
\begin{equation}
\bm{b}^{\,}_{1} \equiv \bm{a}^{\,}_{2} - \bm{a}^{\,}_{3},
\qquad
\bm{b}^{\,}_{2} \equiv \bm{a}^{\,}_{3} - \bm{a}^{\,}_{1},
\qquad
\bm{b}^{\,}_{3} \equiv \bm{a}^{\,}_{1} - \bm{a}^{\,}_{2}.
\end{equation}
The full Bloch Hamiltonian in the first BZ of the triangular sublattice A
inherits a $2\times 2$ sublattice grading that we encode with
the use of the Pauli matrices
$\bm{\tau}=(\tau^{\,}_{x},\tau^{\,}_{y},\tau^{\,}_{z})$.

Following Haldane, we define the
single-particle tight-binding Bloch Hamiltonian
\begin{subequations}
\label{eq:def Haldane Hamiltonian}
\begin{equation}
H(\bm{k})\equiv
H^{\,}_{0}(\bm{k})
+
H^{\,}_{1}(\bm{k})
+
H^{\,}_{2}(\bm{k}).
\label{eq:def Haldane Hamiltonian a}
\end{equation}
The wave vector $\bm{k}$ belongs to the BZ of the triangular sublattice A and
\begin{align}
&
H^{\,}_{0}(\bm{k})\equiv
M\,
\tau^{\,}_{z},
\label{eq:def Haldane Hamiltonian b0}
\\
&
H^{\,}_{1}(\bm{k})\equiv
t^{\,}_{1}
\left(
\sum\limits_{i=1}^{3}
e^{+\mathrm{i} \bm{k} \cdot \bm{a}^{\,}_{i}}
\right)
\frac{\tau^{\,}_{x}+\mathrm{i}\tau^{\,}_{y}}{2}
+
\mathrm{H.c.},
\label{eq:def Haldane Hamiltonian b1}
\\
&
H^{\,}_{2}(\bm{k})\equiv
2\,t^{\,}_{2}
\sum_{i=1}^{3}
\sin (\bm{k} \cdot \bm{b}_i)\,
\tau^{\,}_{z},
\label{eq:def Haldane Hamiltonian b2}
\end{align}
\end{subequations}
where
$M\in\mathbb{R}$ is a staggered chemical potential,
$t^{\,}_{1}>0$
is the amplitude of a uniform nearest-neighbor hopping,
and $ t^{\,}_{2}>0$ is the amplitude of
an imaginary-valued next-nearest-neighbor hopping.
Reversal of time of $H(\bm{k})$ 
is represented by complex conjugation and the substitution
$\bm{k}\to-\bm{k}$. 
The first two terms on the right-hand side of
Eq.~%
(\ref{eq:def Haldane Hamiltonian a})
are even under reversal of time. The last term
on the right-hand side of
Eq.~%
(\ref{eq:def Haldane Hamiltonian a})
is odd under reversal of time.
Hence, the dimensionful coupling $t^{\,}_{2}$ breaks
time-reversal symmetry when nonvanishing.

In the thermodynamic limit,
Hamiltonian (\ref{eq:def Haldane Hamiltonian}) has
two single-particle dispersing bands
\begin{subequations}
\label{eq:Haldane two bands}
\begin{align}
&
H(\bm{k})=
\sum_{\pm}
\varepsilon^{\,}_{\pm}(\bm{k})\,
\left|\pm;\bm{k}\right\rangle
\left\langle\pm;\bm{k}\right|,
\end{align}  
with the dispersions
\begin{align}
&
\varepsilon_{\pm}(\bm{k}) =
\pm\varepsilon(\bm{k}),
\label{eq:Haldane two bands a}
\\
&
\varepsilon(\bm{k}) \equiv
\sqrt{\!
\left[
M\!
+
2 t^{\,}_{2} \sum_{i=1}^{3} \sin (\bm{k} \cdot \bm{b}_i)
\right]^{2}\!\!\!\!
+
t^{2}_{1}
\left|
\sum_{i=1}^{3}
e^{+\mathrm{i} \bm{k} \cdot \bm{a}_i}
\right|^{2}
    }\!\!.
\label{eq:Haldane two bands b}
\end{align}
\end{subequations}
The single-particle spectral symmetry of Hamiltonian
\eqref{eq:def Haldane Hamiltonian}
about the single-particle energy zero is a consequence of
the fact that $H(\bm{k})$,
for some given $\bm{k}$, is odd under  
conjugation by the matrix $\tau^{\,}_{y}$
followed by the transformation
$\bm{a}^{\,}_{1}\mapsto-\bm{a}^{\,}_{1}$,
$\bm{a}^{\,}_{2}\mapsto-\bm{a}^{\,}_{3}$,
and
$\bm{a}^{\,}_{3}\mapsto-\bm{a}^{\,}_{2}$.
In turn, this transformation law 
is nothing but the composition
of $\tau^{\,}_{y}$ acting on the two triangular sublattices
with the reflection about the $y$ axis in the coordinate system
defined by Eq.~%
(\ref{eq:defs a1 a2 a3}), i.e.,
\begin{equation}
\tau^{\,}_{y}\,H(-k^{\,}_{x},k^{\,}_{y})\,\tau^{\,}_{y}=
-H(k^{\,}_{x},k^{\,}_{y}).
\end{equation}
When $M=t^{\,}_{2}=0$, inversion and time-reversal symmetries both hold
simultaneously, the two bands touch at the two nonequivalent corners
\begin{equation}
\bm{K}_{\pm} =
\frac{4\pi}{3\sqrt{3}}
\begin{pmatrix}
\frac{\sqrt{3}}{2}
\\ \\
\pm\frac{1}{2}
\end{pmatrix}
\end{equation}
of the BZ in the close vicinity of which they realize a Dirac spectrum.
Generic values of
$M$ and $t^{\,}_{2}$ break both the inversion and time-reversal symmetries,
while opening a spectral gap at
$\bm{K}_{\pm}$
given by twice the value of
\begin{equation}
m^{\,}_{\pm}(M,t^{\,}_{2}) \equiv \left|M \pm 3 \sqrt{3}\, t^{\,}_{2}\right|.
\end{equation} 
The upper and lower bands have opposite Chern numbers 
\begin{subequations}
\label{eq: Chern numbers}
\begin{equation}
C^{\,}_{\pm}=
\int\limits_{\mathrm{BZ}}\frac{\mathrm{d}^{2}\bm{k}}{2\pi}\,
\Omega^{\,}_{\pm}(\bm{k}),
\label{eq: Chern numbers a}
\end{equation}
where we have introduced the Berry curvature
\begin{align}
\Omega^{\,}_{\pm}(\bm{k})=&\,
\mathrm{i}
\bigg[
\frac{\partial}{\partial k^{\,}_{1}}
\left(
\left\langle\pm;\bm{k}
\left|
\frac{\partial}{\partial k^{\,}_{2}}
\right|\pm;\bm{k}\right\rangle
\right)
\nonumber\\ 
&\,
-
\frac{\partial}{\partial k^{\,}_{2}}
\left(
\left\langle\pm;\bm{k}
\left|
\frac{\partial}{\partial k^{\,}_{1}}
\right|\pm;\bm{k}\right\rangle
\right)
\bigg].
\label{eq: Chern numbers b}
\end{align}
\end{subequations} 
The bands have Chern numbers of unit magnitude when
\begin{equation}
|M|<  \sqrt{3}\,|t^{\,}_{2}|.
\label{eq: condition for nonvanishing Chern numbers}
\end{equation}
They are vanishing otherwise.

\begin{figure}[t]
\centering
\includegraphics[width=\columnwidth]{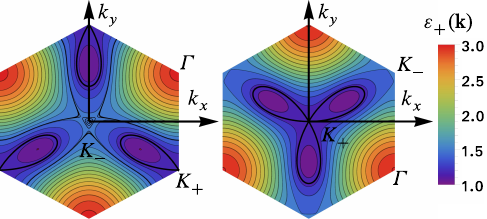}
\caption{
Under appropriate tuning of the staggered chemical potential
[$M=\pm M^{\,}_{0}$ with $M^{\,}_{0}$ defined in Eq.\
(\ref{eq:definition of M0})
with $t^{\,}_{1}=1$ and $t^{\,}_{2}=1/4$],
Hamiltonian (\ref{eq:def Haldane Hamiltonian})  
can be made to host a monkey-saddle singularity at
either one of the $\bm{K}^{\,}_{\pm}$ points. 
While the constant-energy contours of
the monkey-saddle dispersion (\ref{eq: monkey saddle dispersion})
are open, those in the Haldane model in panels (a) and (b)
are closed due to the correction of order $\propto\bm{k}^{4}$
to the monkey-saddle dispersion (\ref{eq: monkey saddle dispersion}).
The monkey saddle with its singular energy contour that
is shaped like the boundary of a three-leaf clover
(bold and black)
is here realized at $\bm{K}^{\,}_{+}$,
while a simple maximum is realized at
$\bm{K}^{\,}_{-}$ higher up in energy.
        }
\label{fig:dispersion}
\end{figure}

We perform the expansion
\begin{subequations}
\label{eq:expansion of Haldane dispersion}
\begin{align}
\varepsilon(\bm{K}^{\,}_{\pm}+\bm{k}) =&\,
m^{\,}_{\pm}(M,t^{\,}_{2})
\nonumber\\
&\,
\mp
9 \sqrt{3}\, t^{\,}_{2}\,
\frac{M \mp M^{\,}_{0}}{4 m^{\,}_{\pm}(M,t^{\,}_{2})}
\,
k^{2}
\nonumber\\
&\,
\pm
\frac{
3 \left[2 t^{2}_{1}\pm \sqrt{3}\, t^{\,}_{2}\, (M \mp M^{\,}_{0}) \right]
     }
     {
8 m^{\,}_{\pm}(M,t^{\,}_{2})
     }
k^{3} \cos(3\theta)
\nonumber\\
&\,
+
\mathcal{O}(k^{4})
\end{align}
of the magnitude (\ref{eq:Haldane two bands b}).
Here, we are using the short-hand notation
\begin{equation}
M^{\,}_{0}\equiv\frac{t^{2}_{1}-18\,t^{2}_{2}}{2\sqrt{3}\, t^{\,}_{2}},
\label{eq:definition of M0}
\end{equation}
at which
\begin{equation}
m^{\,}_{\pm}(M^{\,}_{0},t^{\,}_{2})=
\begin{cases}
\left| \frac{t^{2}_{1}}{2 \sqrt{3}\, t^{\,}_{2}} \right|, & \hbox{if $+$},
\\ \\
\left| \frac{t^{2}_{1} - 36\, t^{2}_{2}}{2 \sqrt{3}\, t^{\,}_{2}} \right|,  & \hbox{if $-$}.
\end{cases}
\end{equation}
\end{subequations}
When the staggered potential takes the value $M=M^{\,}_{0}$,
the magnitude (\ref{eq:Haldane two bands b})
realizes the monkey saddle
\begin{subequations}
\label{eq: bare dispersion MS at K+}
\begin{align}
\varepsilon(\bm{K}^{\,}_{+}+\bm{k}) =&\,
\left|\frac{t^{2}_{1}}{2\sqrt{3}\,t^{\,}_{2}}\right|
+
\frac{3\sqrt{3}}{2}\,|t^{\,}_{2}|\,
k^{3}\,\cos(3\theta)
+\mathcal{O}(k^{4}),
\end{align}
centered about $\bm{K}^{\,}_{+}$
at the energy $m^{\,}_{+}(M^{\,}_{0},t^{\,}_{2})$,
while it realizes the local extremum
\begin{align}
\varepsilon(\bm{K}^{\,}_{-}+\bm{k})=&\,
\left|\frac{t^{2}_{1}-36\,t^{2}_{2}}{2\sqrt{3}\,t^{\,}_{2}}\right|
\nonumber\\
&\,
+
9\sqrt{3}\,|t^{\,}_{2}|
\frac{(t^{2}_{1}-18\,t^{2}_{2}) }{2|t^{2}_{1}-36\,t^{2}_{2}|}\,
k^{2}
\nonumber\\
&\,
-
\frac{3\sqrt{3}\,|t^{\,}_{2}|(t^{2}_{1} + 18\,t^{2}_{2})}
     {4|t^{2}_{1}-36\,t^{2}_{2}|}
\,
k^{3}\,\cos(3\theta)
+\mathcal{O}(k^{4}),
\end{align}
\end{subequations}
centered about $\bm{K}^{\,}_{-}$
at the energy $m^{\,}_{-}(M^{\,}_{0},t^{\,}_{2})$. 
Choosing the value $M=-M^{\,}_{0}$
centers the monkey saddle at
$\bm{K}^{\,}_{-}$ and the local extremum at
$\bm{K}^{\,}_{+}$. In either case, it is always possible
to tune the magnitude of the energy of the monkey saddle
\begin{subequations}
\label{eq:condition for monkey saddle to be at low energies}
\begin{equation}
\mu^{\,}_{\mathrm{ms}}\equiv
\left|\frac{t^{2}_{1}}{2\sqrt{3}\,t^{\,}_{2}}\right|,
\label{eq:condition for monkey saddle to be at low energies a}
\end{equation}
so that it becomes smaller
than the magnitude of the energy of the local extremum
\begin{equation}
\mu^{\,}_{\mathrm{le}}\equiv
\left|\frac{t^{2}_{1}-36\,t^{2}_{2}}{2\sqrt{3}\,t^{\,}_{2}}\right|,
\label{eq:condition for monkey saddle to be at low energies b}
\end{equation}
provided the condition
\begin{equation}
36 (t^{\,}_{2}/t^{\,}_{1})^{2}>2
\ \Longleftrightarrow\
t^{2}_{1}-18\,t^{2}_{2} < 0
\label{eq:condition for monkey saddle to be at low energies c}
\end{equation}
\end{subequations}
holds. Combining condition
(\ref{eq:condition for monkey saddle to be at low energies c})
with the definition of $M^{\,}_{0}$ in
Eq.~%
(\ref{eq:definition of M0})
delivers
\begin{equation}
|M^{\,}_{0}|=\frac{18\,t^{2}_{2}-t^{2}_{1}}{2\sqrt{3}\,|t^{\,}_{2}|}.
\end{equation}
Figure \ref{fig:dispersion}
shows the constant-energy contours of the upper dispersion
of Hamiltonian (\ref{eq:def Haldane Hamiltonian})  
when $M=M^{\,}_{0}$
and $t^{\,}_{2}/t^{\,}_{1}= 1/4$.
The constant-energy contour shaped like a three-leaf clover
is the Fermi surface when the Fermi energy matches the
monkey-saddle energy.

\begin{figure*}[t]
\includegraphics[width=1\textwidth]{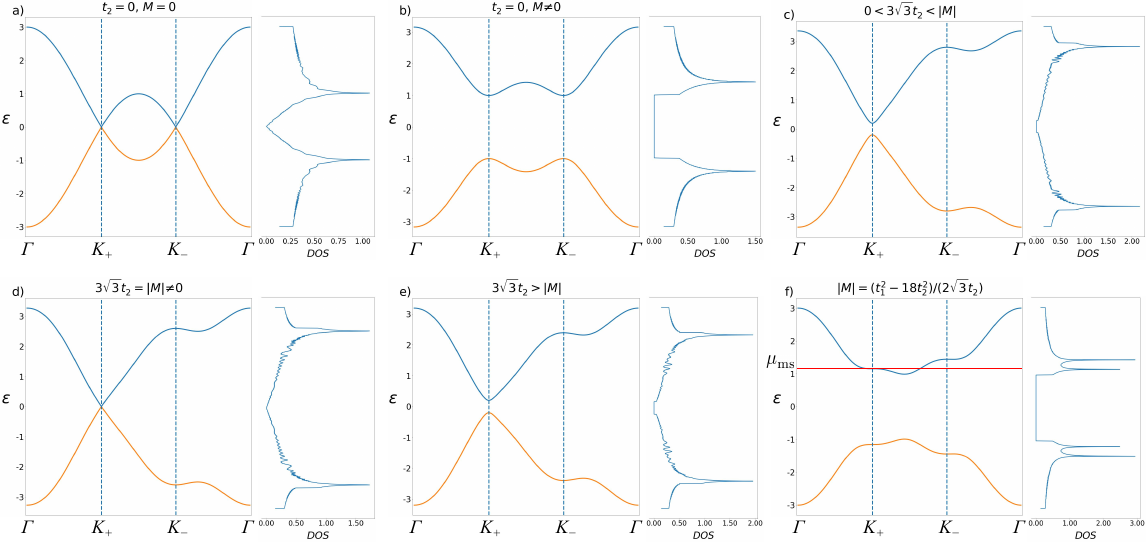}
\caption{
Dispersions (\ref{eq:Haldane two bands a})
and density of states (DOS)
$
\nu(\varepsilon)\:=
N^{-2}
\sum_{\pm}
\sum_{\bm{k}}
\delta\big(\varepsilon-\varepsilon^{\,}_{\pm}(\bm{k})\big)$
for the single-particle tight-binding
Hamiltonian (\ref{eq:def Haldane Hamiltonian}).
The choice $N=1001$ is made and each delta function
entering the DOS is regularized
by a normalized Gaussian of variance
$\sigma^{2}_{\mathrm{Gaussian}}\sim N^{-2}$.
Under tuning of parameters, where we set $t^{\,}_{1}=1$,
Hamiltonian (\ref{eq:def Haldane Hamiltonian})
displays changes in both the band geometry and the band topology. 
We seek to obtain a monkey saddle at
$\bm{K}^{\,}_{+}$
with an extremum at $\bm{K}^{\,}_{-}$, located higher up in energy,
in the upper band of Hamiltonian (\ref{eq:def Haldane Hamiltonian}).
Such a phase is
automatically in the topological regime with
nonvanishing Chern numbers for each filled band. This phase can be reached
starting from the gapless, and time-reversal
invariant Dirac semimetal in (a). As a staggered chemical
potential ($M$) is turned on, a gap at both the $\bm{K}^{\,}_{\pm}$ points
appears as depicted in (b). The dispersions in the neighborhoods of
$\bm{K}^{\,}_{+}$
and
$\bm{K}^{\,}_{-}$
are symmetric because of time-reversal symmetry.
In (c), a small time-reversal breaking next-nearest-neighbor hopping
$t^{\,}_{2}$ that causes an asymmetry between the dispersion around
$\bm{K}^{\,}_{+}$
and that around
$\bm{K}^{\,}_{-}$
is turned on. As the strength of $t^{\,}_{2}$ is increased to
$|M|/3\sqrt{3}$, the gap at $\bm{K}^{\,}_{+}$ closes,
as in (d). By increasing $t^{\,}_{2}$ further as in (e),
the topological regime with nonvanishing Chern number is entered.
Finally, increasing $t^{\,}_{2}$ to satisfy the condition
$M=(t^{2}_{1}-18\,t^{2}_{2})/(2\sqrt{3}\,t^{\,}_{2})$,
we obtain a monkey saddle at $\bm{K}^{\,}_{+}$
and a simple maxima at $\bm{K}^{\,}_{-}$, higher in energy.
This is depicted in (f). The van Hove singularities in the DOS of
panels (a)--(e) have become monkey-saddle singularities at
$\bm{K}^{\,}_{+}$ and van Hove singularities at $\bm{K}^{\,}_{-}$
in panel (f).
        }
\label{fig:evolution_haldane}
\end{figure*}

Assuming that $M$ has been tuned to either $M^{\,}_{0}$ or $-M^{\,}_{0}$,
so as to obtain the monkey saddle at $\bm{K}^{\,}_{+}$ or $\bm{K}^{\,}_{-}$,
respectively, we examine the sign of
$|M^{\,}_{0}|-3\sqrt{3}\,|t^{\,}_{2}|$.
If it is negative, then we are in the regime
for which the Chern number is nonvanishing and the band is
topological. We have
\begin{equation}
|M^{\,}_{0}|-3\sqrt{3}\,|t^{\,}_{2}| =
\frac{18\,t^{2}_{2}-t^{2}_{1}}{2\sqrt{3}\,|t^{\,}_{2}|}-3\sqrt{3}\,|t^{\,}_{2}|=
\frac{-t^{2}_{1}}{2\sqrt{3}\,|t^{\,}_{2}|}< 0.
\end{equation}
Thus, if we tune the chemical potential
to the energy
(\ref{eq:condition for monkey saddle to be at low energies a})
of the monkey saddle and assume that the energy of the local extremum
(\ref{eq:condition for monkey saddle to be at low energies b})
is larger, i.e.,
Eq.~%
(\ref{eq:condition for monkey saddle to be at low energies c})
holds, then the two bands necessarily have nonvanishing Chern numbers.

We plot in Fig.~%
\ref{fig:evolution_haldane}
the two single-particle dispersions
(\ref{eq:Haldane two bands})
along the cuts
$\Gamma-K^{\,}_{+}-K^{\,}_{-}-\Gamma$
in the Brillouin zone
for different values of $M$ and $t^{\,}_{2}$, holding $t^{\,}_{1}$ fixed.
Panel (a) corresponds to the case with two inequivalent Dirac points
at which the upper and lower bands touch.
Panel (b) corresponds to a gap at the two Dirac points
of panel (a) induced by the staggered chemical potential $M$.
Panel (c) shows the effect on panel (b) of a small $t^{\,}_{2}$.
The spectral valley symmetry is broken. 
In panel (d), the competition between $M$ and $t^{\,}_{2}$
results in a gap-closing transition at one of the Dirac points
from panel (a). In panel (e) the gap reopens as
$t^{\,}_{2}$ dominates over $M$. The bands now have the Chern numbers $\pm1$.
In panel (f), $K^{\,}_{+}$ realizes a monkey saddle, while $K^{\,}_{-}$
realizes a local extremum.

We remark that a non-vanishing Hall conductivity results from breaking
time-reversal symmetry. The anomalous Hall conductivity contribution
from the partially filled band varies continuously as a
function of the
band filling. This contribution can be expressed as an integral over
the Brillouin zone of the (regular) Berry curvature over the filled
states. While this integral is continuous
as a function of the chemical potential
as the latter is varied across
the monkey-saddle singularity,
derivatives of the Hall conductivity with respect to the chemical
potential will inherit the singularities in the DOS.

\section{The effects of interactions on a Monkey saddle}
\label{sec: The monkey saddle realizes a non-Fermi liquid}

We consider a two-dimensional gas of spinful electrons
whose single-particle and spin-degenerate dispersion
\begin{equation}
\varepsilon^{\,}_{\mathrm{ms}}(\bm{k})=
-
\varepsilon^{\,}_{\mathrm{ms}}(-\bm{k})
\end{equation}
is the monkey-saddle dispersion defined by Eq.\
(\ref{eq: monkey saddle dispersion}).
The number of energy eigenvalues per unit area in the interval
$(\varepsilon,\varepsilon+\mathrm{d}\varepsilon)$
defines the monkey-saddle density of states
\begin{subequations}
\begin{equation}
\nu^{\,}_{\mathrm{ms}}(\varepsilon):=
\int\frac{\mathrm{d}^{2}\bm{k}}{(2\pi)^{2}}\
\delta
\Big(
\varepsilon
-
\varepsilon(\bm{k})
\Big).
\end{equation}
It is given by\, \cite{monkey_saddle}
\begin{equation}
\nu^{\,}_{\mathrm{ms}}(\varepsilon)=
\frac{1}{2\pi^{3/2}}\,
\frac{\Gamma(1/6)}{\Gamma(2/3)}\,
E^{-2/3}\,
|\varepsilon|^{-1/3}.
\end{equation}
\end{subequations}
As emphasized in Ref.~%
\onlinecite{monkey_saddle},
it displays a power-law singularity at the singular energy
$\varepsilon=0$.

This non-interacting electron gas is perturbed by
a contact density-density interaction for opposite spins.
The quantum dynamics is thus governed by the many-body Hamiltonian
\begin{subequations}
\begin{equation}
\widehat{H}:=
\widehat{H}^{\,}_{\mathrm{ms}}
+
\widehat{H}^{\,}_{\mathrm{int}},
\end{equation}
where the kinetic energy is given by
\begin{equation}
\widehat{H}^{\,}_{\mathrm{ms}}:=
\sum_{\sigma=\uparrow,\downarrow}\
\int\limits_{|\varepsilon(\bm{k})|\leq\Lambda}\mathrm{d}^{2}\bm{k}\
\hat{c}^{\dag}_{\sigma}(\bm{k})
\left[
\varepsilon^{\,}_{\mathrm{ms}}(\bm{k})
-
\mu
\right]
\hat{c}^{\,}_{\sigma}(\bm{k})
\end{equation}
and the interaction is given by
\begin{equation}
\widehat{H}^{\,}_{\mathrm{int}}:=
g\,
\int\mathrm{d}^{2}\bm{r}\,
\hat{c}^{\dag}_{\uparrow}(\bm{r})\,
\hat{c}^{\vphantom{\dag}}_{\uparrow}(\bm{r})\,
\hat{c}^{\dag}_{\downarrow}(\bm{r})\,
\hat{c}^{\vphantom{\dag}}_{\downarrow}(\bm{r}).
\label{eq:spinful RG contact int}
\end{equation}
Here, we have introduced an ultraviolet energy cutoff $\Lambda$,
corresponding to the energy scale at which corrections of order
$\bm{k}^{4}$ in any lattice regularization of
the dispersion (\ref{eq: monkey saddle dispersion}) are comparable to
the $\bm{k}^{3}$ contribution,
$\mu$ denotes the chemical potential,
and $g$ measures the strength of the contact interaction
(a positive $g$ penalizes local double occupancy by electrons). 
The electronic field operators obey fermionic equal-time
anti-commutation relations, i.e., the only
nonvanishing equal-time anti-commutators are
\begin{align}
&
\left\{
\hat{c}^{\vphantom{\dag}}_{\sigma\vphantom{\sigma'}}(\bm{r}),
\hat{c}^{\dag}_{\sigma'}(\bm{r}')
\right\}=
\delta^{\,}_{\sigma,\sigma'}\,
\delta(\bm{r}-\bm{r}'),
\\
&
\left\{
\hat{c}^{\vphantom{\dag}}_{\sigma\vphantom{\sigma'}}(\bm{k}),
\hat{c}^{\dag}_{\sigma'}(\bm{k}')
\right\}=
\delta^{\,}_{\sigma,\sigma'}\,
\delta(\bm{k}-\bm{k}').
\end{align}
\end{subequations}
The chemical potential is fixed by the number
$N^{\,}_{\mathrm{e}}$ of electrons in the large area $A$.
Henceforth, we set the units such that
\begin{subequations}
\begin{equation}
\hbar=1,
\quad
k^{\,}_{\mathrm{B}}=1,
\end{equation}
for the Planck and Boltzmann constants, respectively.
In these units, temperature $T$ has units of energy and
time has units of inverse energy.
The grand-canonical partition function at the inverse temperature
$\beta=1/T$ is 
\begin{equation}
Z(\beta,\mu):=
\mathrm{Tr}\, e^{-\beta \widehat{H}},
\qquad
N^{\,}_{\mathrm{e}}=
\beta^{-1}\,
\left(\frac{\partial\,\ln Z}{\partial\,\mu}\right)(\beta,\mu).
\end{equation}
\end{subequations}

The decay rate $\Gamma(g,T)$
of quasi-particles when $\mu=0$
arising from the contact interaction
was calculated in Ref.~%
\onlinecite{monkey_saddle}
to the first non-trivial order in perturbation theory.
It is given by
\begin{equation}
\Gamma(g,T)=
C\
g^{2}\
\nu^{2}(T)\,
T
\sim
T^{1/3}
\label{eq: decay rate to lowest order in perturbation theory dimfull}
\end{equation}
with $C$ a positive numerical constant
(that is calculated in the limit $\Lambda\to\infty$).
For comparison, the decay rate of a Fermi liquid in two-dimensional space
scales with temperature as $T^{2}$ up to a multiplicative
logarithmic correction. 
However, this non-Fermi-liquid decay rate does not hold all the way
to vanishing temperature as higher-order corrections in perturbation theory
in powers of $g$ acquire power-law corrections
in the temperature with negative scaling exponents,
since the dimensionless expansion parameter is
$g\,\nu(T)$.

Renormalization-group techniques can be useful when perturbation theory
is not converging uniformly.
After tracing over all electrons whose energies are within the energy shell
$
\Lambda-\mathrm{d}\Lambda\leq
|\varepsilon^{\,}_{\mathrm{ms}}(\bm{k})|\leq
\Lambda
$
with
\begin{equation}
\frac{\mathrm{d}\Lambda}{\Lambda}=\mathrm{d}\ell,
\end{equation}
infinitesimal, it is possible to preserve the form invariance of the
grand-canonical partition function provided the dimensionless
temperature
\begin{subequations}
\begin{equation}
\overline{T}:=\frac{T}{\Lambda},
\end{equation}
the dimensionless chemical potential 
\begin{equation}
\overline{\mu}:=\frac{\mu}{\Lambda},
\end{equation}
and the dimensionless interaction strength
\begin{equation}
\overline{g}:=
\nu(\Lambda)\,g,
\end{equation}
\end{subequations}
obey the renormalization-group (RG) equations
\begin{subequations}
\label{eq: RG flows when dispersion is odd under kto-k}
\begin{align}
&
\frac{\mathrm{d}\,\overline{T}}{\mathrm{d}\,\ell}=
\overline{T},
\\
&
\frac{\mathrm{d}\,\overline{\mu}}{\mathrm{d}\,\ell}=
\left[
1
-
\frac{
\overline{g}
    }
     {
2\overline{T}\,
\cosh^{2}\left(\frac{1}{2\overline{T}}\right)
     }
\right]
\overline{\mu},
\\
&
\frac{\mathrm{d}\,\overline{g}}{\mathrm{d}\,\ell}=
\frac{1}{3}\,\overline{g}.
\end{align}
\end{subequations}
These RG equations were derived perturbatively 
about the fixed point
\begin{equation}
\overline{T}^{\,\star}=\overline{\mu}^{\,\star}=\overline{g}^{\,\star}=0
\label{eq: trivial fixed point}  
\end{equation}
up to order $\bar{g}^{3}$
in Refs.\
\onlinecite{monkey_saddle}
and
\onlinecite{supermetal}.
Whereas $\overline{T}$ and $\overline{g}$ flow to strong coupling,
i.e., beyond the range of validity of these perturbative RG
flows, the beta function of the dimensionless chemical potential
$\overline{\mu}$
undergoes a sign change if and only if the initial value of
$\overline{g}$ is larger than the initial value of
$
2\overline{T}\,
\cosh^{2}\left(\frac{1}{2\overline{T}}\right)
$.
If the initial conditions correspond to vanishing temperature,
the RG equations
(\ref{eq: RG flows when dispersion is odd under kto-k})
simplify to
\begin{subequations}
\begin{equation}
\frac{\mathrm{d}\,\overline{\mu}}{\mathrm{d}\,\ell}=
\overline{\mu},
\qquad
\frac{\mathrm{d}\,\overline{g}}{\mathrm{d}\,\ell}=
\frac{1}{3}\,\overline{g}.
\end{equation}
If the initial conditions correspond to vanishing chemical potential
the RG equations
(\ref{eq: RG flows when dispersion is odd under kto-k})
simplify to
\begin{equation}
\frac{\mathrm{d}\,\overline{T}}{\mathrm{d}\,\ell}=
\overline{T},
\qquad
\frac{\mathrm{d}\,\overline{g}}{\mathrm{d}\,\ell}=
\frac{1}{3}\,\overline{g}.
\end{equation}
\end{subequations}

One possible interpretation of this RG flow to strong coupling
is a Stoner instability to an itinerant ferromagnetic phase,
as can be confirmed by a mean-field analysis\, \cite{supermetal}.
Pomeranchuk instabilities
(area-preserving deformations of the
three-leaf clover Fermi surface into either a single Fermi surface
enclosing the monkey-saddle singularity at $\bm{K}^{\,}_{+}$, say,
or three disconnected Fermi surfaces
that do not enclose the monkey-saddle singularity)
are also possible.
Any superconducting instability must be of the
Fulde-Ferrell-Larkin-Ovchinnikov (FFLO) type
with the characteristic monkey-saddle wave vector $\bm{K}^{\,}_{+}$, say.
More exotic instabilities such as a fractional Chern
insulator when the band hosting the monkey saddle has a nonvanishing
Chern number, the filling fraction is 1/3 at the monkey saddle,
and the interaction strength is larger than the band width, say,
cannot be ruled out owing to the DOS at the monkey saddle.
Nonperturbative techniques are needed to establish
the fate of the monkey saddle when perturbed by a contact interaction.

We are going to use the mean-field approximation to argue that
the monkey-saddle singularity is unstable when we
elevate the spinless fermions in Hamiltonian
(\ref{eq:def Haldane Hamiltonian})
to electrons with spin-1/2 and add
an on-site repulsive Hubbard interaction with coupling $U>0$
and a next-nearest-neighbor repulsive interaction  with coupling $V>0$.

For the case of a two-dimensional gas of spinless electrons, 
there is no quartic density-density 
contact interaction as in Eq.\ \eqref{eq:spinful RG contact int}. 
The lowest-order interaction term that we may add 
is
\begin{equation}
\propto[\hat{c}^{\dagger}(\bm{r})\,\bm{\nabla}\hat{c}(\bm{r})]^{2}.
\end{equation}
This interaction is irrelevant by power counting
and is thus not expected to destabilize the monkey saddle for
small values of its coupling. Accordingly,
we are going to show that a monkey-saddle singularity
can be stabilized by fine-tuning lattice parameters
in the presence of repulsive nearest-neighbor interactions 
within a mean-field approximation.

\section{Mean-field analysis}
\label{sec: Mean field analysis}

In this section, we analyze the stability of the monkey-saddle
singularity in the spectrum of the Hamiltonian
\eqref{eq:def Haldane Hamiltonian}
against short-range interactions at the mean-field
level.  We treat the cases of spinful and spinless electrons
separately.  For the former case, we consider repulsive on-site
Hubbard and nearest-neighbor interactions.  For the latter case, we
only consider a repulsive nearest-neighbor interaction.

\subsection{Spinful case}
\label{subsec:MF Spinful case}

We presume Hamiltonian (\ref{eq:def Haldane Hamiltonian}) for spinful
electrons fine-tuned to a monkey saddle located at $\bm{K}^{\,}_{+}$
in the upper ($+$) band that is perturbed by a repulsive on-site Hubbard
interaction of strength $U$ and a repulsive nearest-neighbor interaction of
strength $V$, given by
\begin{subequations}
\label{eq:interaction Hamiltonians}
\begin{align}
&
\widehat{H}^{\,}_{U}
=
U
\sum_{\bm{r}\in \Lambda}
\Big(
\hat{n}^{\,}_{\mathrm{A},\uparrow,\bm{r}}\,
\hat{n}^{\,}_{\mathrm{A},\downarrow,\bm{r}}
+
\hat{n}^{\,}_{\mathrm{B},\uparrow,\bm{r}+\bm{a}^{\,}_{1}}\,
\hat{n}^{\,}_{\mathrm{B},\downarrow,\bm{r}+\bm{a}^{\,}_{1}}
\Big),
\label{eq:interaction Hamiltonians a}
\\
&
\widehat{H}^{\,}_{V}
=
V
\sum_{\bm{r}\in \Lambda}
\sum_{i=1}^{3}
\hat{n}^{\,}_{\mathrm{A},\bm{r}}\,
\hat{n}^{\,}_{\mathrm{B},\bm{r}+\bm{a}^{\,}_{i}},
\label{eq:interaction Hamiltonians b}
\end{align}
\end{subequations}
respectively. 
Here, we denote with $\Lambda$ the triangular
Bravais lattice hosting the A sites. The honeycomb lattice is
made of $|\Lambda|$ unit cells, each one containing two sites
labeled by A and B.
The total number of sites in the honeycomb lattice is thus $2|\Lambda|$.
Hereby, we have introduced the spin 
and position resolved fermion number 
operators
\begin{subequations}
\begin{align}
&
\hat{n}^{\,}_{\mathrm{A},\sigma,\bm{r}}
=
\hat{c}^{\dagger}_{\mathrm{A},\sigma,\bm{r}}
\hat{c}^{\,}_{\mathrm{A},\sigma,\bm{r}},
\quad
\hat{n}^{\,}_{\mathrm{B},\sigma,\bm{r}+\bm{a}^{\,}_{i}}
=
\hat{c}^{\dagger}_{\mathrm{B},\sigma,\bm{r}+\bm{a}^{\,}_{i}}
\hat{c}^{\,}_{\mathrm{B},\sigma,\bm{r}+\bm{a}^{\,}_{i}},
\\
&
\hat{n}^{\,}_{\mathrm{A},\bm{r}}
=
\sum_{\sigma=\pm}
\hat{n}^{\,}_{\mathrm{A},\sigma,\bm{r}},
\quad
\hat{n}^{\,}_{\mathrm{B},\bm{r}+\bm{a}^{\,}_{i}}
=
\sum_{\sigma=\pm}
\hat{n}^{\,}_{\mathrm{B},\sigma,\bm{r}+\bm{a}^{\,}_{i}},
\end{align}
where $\hat{c}^{\dag}_{\mathrm{A},\sigma,\bm{r}}$
and
$\hat{c}^{\dag}_{\mathrm{B},\sigma,\bm{r}+\bm{a}^{\,}_{i}}$
create an electron with spin $\sigma$ on the A and B sublattices 
at positions $\bm{r}$ and $\bm{r}+\bm{a}^{\,}_{i}$, respectively.
\end{subequations}

We employ five mean-field order parameters:
the uniform charge density $n^{\,}_{\mathrm{e}}$,
the uniform magnetization density $\overline{m}$, and 
the three uniform, directed, nearest-neighbor bond density 
order parameters $\overline{\chi}^{\,}_{i}$.
These five order parameters are defined as the ground-state expectation values 
of the local operators
\begin{subequations}
\label{eq:order parameter operators}
\begin{align}
&
\hat{n}^{\,}_{\bm{r}}
=
\hat{n}^{\,}_{\mathrm{A},\bm{r}}
+
\hat{n}^{\,}_{\mathrm{B},\bm{r}+\bm{a}^{\,}_{1}},
\label{eq:order parameter operators a}
\\
&
\hat{m}^{\,}_{\bm{r}}
=
\sum_{\sigma=\pm}
\sigma\,
\left(
\hat{n}^{\,}_{\mathrm{A},\sigma,\bm{r}}
+
\hat{n}^{\,}_{\mathrm{B},\sigma,\bm{r}+\bm{a}^{\,}_{1}}
\right),
\label{eq:order parameter operators b}
\\
&
\hat{\chi}^{\,}_{i,\sigma,\sigma',\bm{r}}
=
\hat{c}^{\dagger}_{\mathrm{A},\sigma,\bm{r}}\,
\hat{c}^{\,}_{\mathrm{B},\sigma',\bm{r}+\bm{a}^{\,}_{i}}
+
\mathrm{H.c.},
\label{eq:order parameter operators c}
\end{align}
\end{subequations}
respectively.
We make the mean-field Ansatz 
\begin{subequations}
\label{eq:MF Ansatz}
\begin{align}
&
\left\langle
\hat{n}^{\,}_{\bm{r}}\,
\right\rangle
=
n^{\,}_{\mathrm{e}},
\label{eq:MF Ansatz a}
\\
&
\left\langle
\hat{m}^{\,}_{\bm{r}}\,
\right\rangle
=
\overline{m},
\label{eq:MF Ansatz b}
\\
&
\left\langle
\hat{\chi}^{\,}_{i,\sigma,\sigma',\bm{r}}\,
\right\rangle
=
\delta^{\,}_{\sigma\,\sigma'}\,
\left(
\overline{\chi}
+
\delta^{\,}_{i,1}\,
\overline{\chi}^{\,}_{1}
\right),
\label{eq:MF Ansatz c}
\end{align}
\end{subequations}
where $\left\langle\cdots\right\rangle$ denotes the 
expectation value over the mean-field ground state.
In the mean-field Ansatz \eqref{eq:MF Ansatz},
we assume that the order parameters are independent of 
the position $\bm{r}$, i.e., the Ansatz \eqref{eq:MF Ansatz}
does not include charge-density, spin-density, or bond-density waves.
This assumption is justified since (i) the single-particle energies at
$\bm{K}^{\,}_{+}$ and $\bm{K}^{\,}_{-}$
are separated in energy (ii) and there are no momentum-conserving 
nesting vectors that connect two points from the Fermi surface
when the chemical potential is tuned close to the 
monkey-saddle energy.
Consequently, there is no band folding in the Brillouin zone and 
the mean-field ground state remains metallic for any noninteger filling fraction.

The mean-field Ansatz (\ref{eq:MF Ansatz a}) for the charge density
fixes the chemical potential such that
the filling fraction of the interacting system
coincides with that of the noninteracting model.
The mean-field Ansatz \eqref{eq:MF Ansatz b} assumes a
ferromagnetic ground state whenever $|\overline{m}|>0$ for which
the spin-rotation symmetry is spontaneously broken [time-reversal symmetry 
is explicitly broken in the Hamiltonian \eqref{eq:def Haldane Hamiltonian} by the 
next-nearest neighbor hopping term \eqref{eq:def Haldane Hamiltonian b2}].
The mean-field Ansatz \eqref{eq:MF Ansatz c} assumes 
a uniform bond-density order parameter
$\overline{\chi}$ that does not break the 
$\mathbb{Z}^{\,}_{3}$-rotation symmetry 
that is modulated by $\overline{\chi}^{\,}_{1}$ along the $\bm{a}^{\,}_{1}$ direction.
Any nonvanishing $|\overline{\chi}^{\,}_{1}|$
breaks the $\mathbb{Z}^{\,}_{3}$-rotation symmetry spontaneously,
while preserving the reflection symmetry 
along the $\bm{a}^{\,}_{1}$ direction.

\begin{figure*}[t]
\centering
\includegraphics[width=\textwidth]{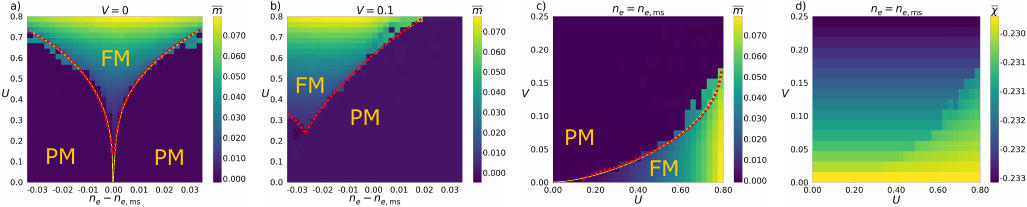}
\caption{
The mean-field phase diagram at zero temperature
in the coupling space spanned by
the filling fraction $n^{\,}_{e}$,
the on-site repulsive interaction $U$,
and
the nearest-neighbor repulsive interaction $V$
is obtained from solving numerically the
mean-field equations
\eqref{eq:MF equations op formalism}
(All energies are measured in units of $t^{\,}_{1}$.).
Dashed red lines show the approximate phase boundaries
in panels (a), (b), and (c).
The yellow solid lines shows the approximate phase boundaries
in the thermodynamic limit in panels (a) and (c).
(a)
Two-dimensional cut for the values taken by $\overline{m}$
when $V=0$ in units of $t^{\,}_{1}$.
A Stoner instability towards itinerant ferromagnetism 
takes place for any nonvanishing Hubbard interaction
$U>U^{\,}_{\mathrm{c}}$ for given $n^{\,}_{e}$.
The minimum value $U^{\,}_{\mathrm{c},\mathrm{min}}$
taken by $U^{\,}_{\mathrm{c}}$ occurs for
$n^{\,}_{e}=n^{\,}_{e,\mathrm{ms}}$.
The nonvanishing value of $U^{\,}_{\mathrm{c},\mathrm{min}}$
above which ferromagnetism is established
when the filling fraction is fine tuned to the monkey saddle, i.e.,
$n^{\,}_{e}=n^{\,}_{e,\mathrm{ms}}$,
is due to a finite-size effect that cuts off the diverging DOS.
In the thermodynamic limit,
$U^{\,}_{\mathrm{c},\mathrm{min}}\to0$
at $n^{\,}_{e}=n^{\,}_{e,\mathrm{ms}}$.
(b)
Two-dimensional cut for the values taken by $\overline{m}$
when $V=0.1$ in units of $t^{\,}_{1}$.
A nonvanishing $V>0$ has two effects. It increases
the minimum value of $U^{\,}_{\mathrm{c}}$ from panel (a) to a
value that remains nonvanishing in the thermodynamic limit.
The position of the minimum value of $U^{\,}_{\mathrm{c}}$ from panel (a)
is shifted along the $n^{\,}_{e}$ axis.
(c)
Two-dimensional cut for the values taken by $\overline{m}$
when $n^{\,}_{e}=n^{\,}_{e,\mathrm{ms}}$.
The critical value $U^{\,}_{\mathrm{c}}$ above which ferromagnetism
sets in is an increasing function of $V$.
(d) Two-dimensional cut for the values taken by $\overline{\chi}$
when $n^{\,}_{e}=n^{\,}_{e,\mathrm{ms}}$.
The value of $\overline{\chi}$ is nonvanishing everywhere.
}
\label{fig:MF phase diag}
\end{figure*}

After performing the mean-field approximation,
the dispersions (\ref{eq:Haldane two bands})
become
\begin{subequations}
\label{eq:MF dispersion}
\begin{equation}
\overline{\varepsilon}^{\,}_{\tau,\sigma,\bm{k}}=
\tau\,\overline{\varepsilon}^{\,}_{\bm{k}}
-\frac{1}{2}\sigma\,U\,\overline{m},
\label{eq:MF dispersion a}
\end{equation}
with
\begin{align}
&
\overline{\varepsilon}^{\,}_{\bm{k}}=
\sqrt{
M^{2}_{\bm{k}}
+
\left|
\Phi^{\,}_{\bm{k}}
\right|^{2}
},
\label{eq:MF dispersion b}
\\
&
M^{\,}_{\bm{k}}=
M
+
2\,t^{\,}_{2}
\sum_{i=1}^{3}
\sin (\bm{k} \cdot \bm{b}_i),
\label{eq:MF dispersion c}
\\
&
\Phi^{\,}_{\bm{k}}
=
\sum\limits_{i=1}^{3}
\overline{t}^{\,}_{1,i}\,
e^{+\mathrm{i}\bm{k}\cdot\bm{a}^{\,}_{i}},
\label{eq:MF dispersion d}
\\
&
\overline{t}^{\,}_{1,i}=
\begin{cases}
t^{\,}_{1}
-
V\,\left(\overline{\chi}+
\overline{\chi}^{\,}_{1}\right),&
i=1,
\\&\\
t^{\,}_{1}-V\,\overline{\chi},&
i=2,3,
\end{cases}
\label{eq:MF dispersion e}
\end{align}
\end{subequations}
where $\tau = \pm$ is the band index and $\sigma = \pm$ is the spin
index. The self-consistent mean-field equations
corresponding to the Ansatz \eqref{eq:MF Ansatz} are
\begin{subequations}
\label{eq:MF equations op formalism}
\begin{align}
&
n^{\,}_{e}
=
\frac{1}{|\Lambda|}
\sum_{\bm{k}, \tau, \sigma}
f^{\,}_{\mathrm{FD}}
\left(
\overline{\varepsilon}^{\,}_{\tau,\sigma,\bm{k}}
-
\overline{\mu}
\right),
\label{eq:MF equations op formalism a}
\\
&
\overline{m}
=
\frac{1}{|\Lambda|}
\sum_{\bm{k}, \tau, \sigma}
\sigma\,
f^{\,}_{\mathrm{FD}}
\left(
\overline{\varepsilon}^{\,}_{\tau,\sigma,\bm{k}}
-
\overline{\mu}
\right),
\label{eq:MF equations op formalism b}
\\
&
\overline{\chi}
=
\frac{1}{2|\Lambda|}
\sum_{\bm{k}, \tau, \sigma}
\tau
\mathrm{Re}
\left\{
e^{\mathrm{i}\bm{k}\cdot\bm{a}^{\,}_{2}}\,
\frac{\Phi^{*}_{\bm{k}}}{2\,\overline{\varepsilon}^{\,}_{\bm{k}}}
\right\}
f^{\,}_{\mathrm{FD}}
\left(
\overline{\varepsilon}^{\,}_{\tau,\sigma,\bm{k}}
-
\overline{\mu}
\right),
\label{eq:MF equations op formalism c}
\\
&
\overline{\chi}^{\,}_{1}
=
\frac{1}{2|\Lambda|}
\sum_{\bm{k}, \tau, \sigma}
\tau
\mathrm{Re}
\left\{
e^{\mathrm{i}\bm{k}\cdot\bm{a}^{\,}_{1}}
\frac{\Phi^{*}_{\bm{k}}}{2\,\overline{\varepsilon}^{\,}_{\bm{k}}}
\right\}
f^{\,}_{\mathrm{FD}}
\left(
\overline{\varepsilon}^{\,}_{\tau,\sigma,\bm{k}}
-
\overline{\mu}
\right)
-
\overline{\chi}.
\label{eq:MF equations op formalism d}
\end{align}
\end{subequations}
Here, the Fermi-Dirac distribution is
\begin{equation}
f^{\,}_{\mathrm{FD}}(\epsilon):=
\begin{cases}
\frac{1}{e^{\epsilon/T} + 1}, & T > 0
\\
&
\\
\Theta(-\epsilon), & T = 0,
\end{cases}
\end{equation}
where $T$ is the temperature (in the units with the Boltzmann constant
set to unity), and $\Theta (x)$ is the step function equal to $1$ for
positive $x$ and $0$ otherwise.  The chemical potential
$\overline{\mu}$ is determined by solving self-consistency
equation \eqref{eq:MF equations op formalism a} where the charge density
$n^{\,}_{e}$ is that of the noninteracting Hamiltonian
\eqref{eq:def Haldane Hamiltonian}.
We denote by $\mu^{\,}_{e}$ the chemical potential
that delivers the charge density $n^{\,}_{e}$ for the noninteracting
dispersion. We note that mean-field Ansatz \eqref{eq:MF Ansatz c}
assumes that the bond-density order parameter is the same for the
$\bm{a}^{\,}_{2}$- and $\bm{a}^{\,}_{3}$-directions.  Therefore, in
the self-consistency equations \eqref{eq:MF equations op formalism c}
and \eqref{eq:MF equations op formalism d}, we could have equivalently
chosen $\bm{a}^{\,}_{3}$ instead of $\bm{a}^{\,}_{2}$.  One can also
generalize Ansatz \eqref{eq:MF Ansatz c} by introducing three separate
bond-density order parameters, one for each direction
$\bm{a}^{\,}_{i}$.  Such a more general mean-field Ansatz, while being
computationally heavier, does not change our results within the
investigated parameter range.

The self-consistent mean-field equation
\eqref{eq:MF equations op formalism}
consists of four unknowns, 
$\left\{
\overline{\mu},\,\overline{m},\,\overline{\chi},\,\overline{\chi}^{\,}_{1}
\right\}$
that are to be determined as a function of three parameters 
$\left\{n^{\,}_{e},\,U,\,V\right\}$. 
We have solved Eqs.~%
\eqref{eq:MF equations op formalism} numerically 
on the Brillouin zone $\Omega^{\,}_{\mathrm{BZ}}$ discretized on a $501\times 501$
grid of $\bm{k}$-points. All energy scales are measured in units of $t^{\,}_{1}$.
We have set $t^{\,}_{2}=0.25$ and $M=M^{\,}_{0}$
for which a monkey-saddle singularity appears. We consider repulsive couplings
$U\geq0$ and $V\geq0$. The coupling $V$ is taken to be 
smaller than the energy difference in the upper band
between the monkey saddle at $\bm{K}^{\,}_{+}$
and the local extremum $\mu^{\,}_{\mathrm{le}}$ at $\bm{K}^{\,}_{-}$.
For our choice of parameters,
this difference is $\Delta\mu =
\mu^{\,}_{\mathrm{le}}-\mu^{\,}_{\mathrm{ms}}\approx 0.29$ in units of $t^{\,}_{1}$. 
For interaction strengths larger than $\Delta \mu$, a bond-density wave with a
nonzero wave vector is a potential instability 
that is not contained in the Ansatz \eqref{eq:MF Ansatz}.

In Fig.~%
\ref{fig:MF phase diag}, the mean-field solutions for the
order parameters $\overline{m}$ and $\overline{\chi}^{\,}$ are
shown as functions of the parameters
$n^{\,}_{e}$, $U$, and $V$.  We only
find two phases, an itinerant phase supporting ferromagnetism
($\overline{m}\neq 0$) and an itinerant phase that is paramagnetic
($\overline{m}= 0$).  The phase boundaries are shown by red dashed
lines.  Within the parameter space of interest, we do not find
the signature of a Pomeranchuk instability ($\overline{\chi}^{\,}_{1}\neq0$)
that would break spontaneously the lattice
$\mathbb{Z}^{\,}_{3}$-rotation symmetry.
Nevertheless, the monkey-saddle singularity is unstable against any
finite repulsive, nearest-neighbor interaction $V$ as we
shall explain shortly.

\begin{figure}[t]
\centering
\includegraphics[width=0.95\columnwidth]{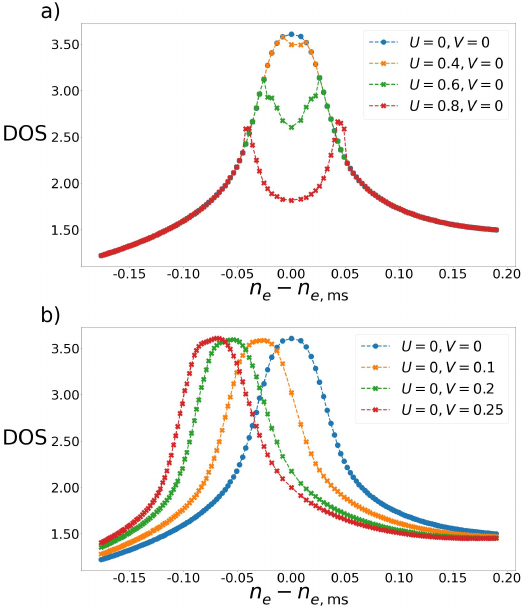}
\caption{
The regularized mean-field DOS
as a function of the deviation
$n^{\,}_{\mathrm{e}}-n^{\,}_{\mathrm{e,ms}}$
of the filling fraction $n^{\,}_{\mathrm{e}}$
from the monkey-saddle filling fraction
$n^{\,}_{\mathrm{e,ms}}$
is plotted for different values of the interaction strengths
$U\geq0$ and $V\geq0$.
The $\delta$-functions in the mean-field DOS are regularized by normalized
Gaussians of variance $\sigma^{2}_{\mathrm{Gaussian}}\sim 10/N^{2}$ with $N=501$.
In panel (a), $U$ is increased holding $V=0$.
The single regularized peak in the noninteracting DOS
is split into two peaks by a nonvanishing $U$. Contrary to the height
of the monkey-saddle peak of the regularized noninteracting DOS,
the height of these secondary peaks remains finite
in the limit $\sigma^{\,}_{\mathrm{Gaussian}}\to0$.
In panel (b), $V$ is increased holding $U=0$.
The single peak in the regularized noninteracting DOS is translated to
the left by a nonvanishing $V$. This observation can be explained by
$V>0$ inducing quadratic perturbations to the monkey-saddle dispersion
(\ref{eq: monkey saddle dispersion})
at the mean-field level. These quadratric perturbations turn
the monkey-saddle singularity into a central local extremum surrounded by
three van Hove saddle singularities.
       }
\label{fig:int effect on DOS}
\end{figure}

For $V=0$, we find that the Stoner instability destroys the
monkey-saddle singularity for any repulsive Hubbard interaction
strength $U>0$ when the filling fraction is tuned to be at the monkey
saddle ($n^{\,}_{e}=n^{\,}_{e,\mathrm{ms}}$).  
This is signaled by (i) the nonvanishing 
magnetization density $\overline{m}$ in
Fig.~%
\ref{fig:MF phase diag}(a)
for $U\geq U^{\,}_{\mathrm{c}}$
where for any finite lattice size $|\Lambda|$
the critical interaction strength $U^{\,}_{\mathrm{c}}$
is minimized as a function of
$n^{\,}_{e}$ when $n^{\,}_{e}=n^{\,}_{e,\mathrm{ms}}$
(ii) whereby we have verified that this minimum
$U^{\,}_{\mathrm{c},\mathrm{min}}$ of
$U^{\,}_{\mathrm{c}}$ decreases with increasing lattice size $|\Lambda|$
with the extrapolated limit $U^{\,}_{\mathrm{c},\mathrm{min}}\to0$ as
$|\Lambda|\to\infty$.
This mean-field calculation thus
confirms the intuition based on Sec.\
\ref{sec: The monkey saddle realizes a non-Fermi liquid}
that the flow of the on-site
interaction to strong coupling is a diagnostic of a Stoner instability
(an itinerant Fermi-liquid phase supporting ferromagnetic long-range order)
as opposed to a featureless (without any long-range order)
non-Fermi-liquid phase. The corresponding effect on the mean-field
DOS is shown in Fig.\  \ref{fig:int effect on DOS}(a).
The mean-field treatment of the on-site repulsive interaction
only changes the spin-resolved chemical potentials.
This will not affect the non-interacting DOS at values of
$n^{\,}_{\mathrm{e}}-n^{\,}_{\mathrm{e,ms}}$
for which the non-interacting DOS is too small to
induce a Stoner instability. However, a Stoner instability must happen
close enough to the monkey-saddle filling fraction $n^{\,}_{\mathrm{e,ms}}$
for any non-vanishing value of $U>0$,
thereby cutting off the monkey-saddle divergence of the noninteracting DOS
at the monkey-saddle filling fraction.
Correspondingly, the regularized mean-field DOS shows
the double-peak shape from Fig.\ \ref{fig:int effect on DOS}(a).

Turning on a nonvanishing $V$ has two effects shown in
Figs.~%
\ref{fig:MF phase diag}(b) and~%
\ref{fig:MF phase diag}(c).
First, the value of $U^{\,}_{\mathrm{c}}$ above which ferromagnetism takes place
is larger for $V=0.1$ than for $V=0$ (measured in units of $t^{\,}_{1}$)
and this value remains nonvanishing in the thermodynamic limit.
Second, the minimal value $U^{\,}_{\mathrm{c},\mathrm{min}}$ of $U^{\,}_{\mathrm{c}}$
in Fig.~%
\ref{fig:MF phase diag}(b) is found at a 
filling fraction $n^{\,}_{e,\mathrm{min}}<n^{\,}_{e,\mathrm{ms}}$.
Both effects can be understood as the renormalization
\eqref{eq:MF dispersion e} of the nearest-neighbor hopping amplitude
$t^{\,}_{1}$ for any nonvanishing $V$. Indeed,
the uniform bond density $\overline{\chi}$ defined in Eq.\
\eqref{eq:MF equations op formalism c} is nonvanishing for any
interaction strengths $U$ and $V$, and for any filling fraction except
for the completely filled ($n^{\,}_{e}=4$) or completely empty
($n^{\,}_{e}=0$) bands.  Any finite interaction strength $V$ thus
results in corrections proportional to $k^{2}$ in the expansion
\eqref{eq:expansion of Haldane dispersion} that
had been set to $0$ by fine tuning the value of the
staggered chemical potential $M$ to $M^{\,}_{0}$
so as to obtain the bare monkey-saddle dispersion
(\ref{eq: bare dispersion MS at K+}).
Under the $k^{2}$-perturbation, the monkey-saddle singularity 
turns into  a central local extremum surrounded by
three van Hove saddle singularities with dispersions 
$\sim k^{2}_{x}-k^{2}_{y}$~\cite{monkey_saddle}.
Consequently, the monkey-saddle singularity disappears through a
Lifshitz transition by which the topology of the Fermi surface changes.
This renormalization has two effects.
It moves the position of the maximum of the mean-field DOS
(i.e., the position of the minimum $U^{\,}_{\mathrm{c},\mathrm{min}}$)
to a value
$n^{\,}_{e,\mathrm{min}}<n^{\,}_{e,\mathrm{ms}}$
[see Fig.\  \ref{fig:int effect on DOS}(b)].
It regularizes the diverging monkey-saddle DOS 
to a large but finite value at the filling fraction
$n^{\,}_{e}=n^{\,}_{e,\mathrm{ms}}$
[see Fig.\  \ref{fig:int effect on DOS}(b)].
Figure\ \ref{fig:MF phase diag}(c)
shows the suppression of the critical interaction strength
$U^{\,}_{\mathrm{c}}$ at the monkey-saddle singularity
with increasing $V$.
Figure\ \ref{fig:MF phase diag}(d)
demonstrates that the uniform bond-density $\overline{\chi}$
is nonvanishing in the same field of view as in Fig.~%
\ref{fig:MF phase diag}(c). In contrast, the
non-isotropic bond-density $\overline{\chi}^{\,}_{1}$
is found to  be vanishing everywhere in coupling space within
the numerical error bars. In other words, we did not
find any evidence for a Pomeranchuk instability.

\begin{figure}[t]
\centering
\includegraphics[width=\columnwidth]{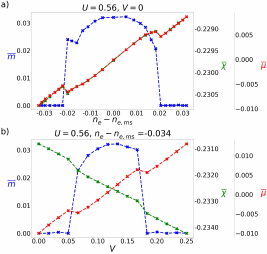}
\caption{
Dependencies of the uniform magnetization $\overline{m}$,
uniform bond-density $\overline{\chi}$,
and chemical potential $\overline{\mu}$
along one-dimensional cuts
in coupling space at zero temperature
as is explained in the text.
}
\label{fig:MF transitions}
\end{figure}

Figure \ref{fig:MF transitions}
shows the variations of the uniform magnetization $\overline{m}$,
the uniform bond-density wave $\overline{\chi}$, and the
chemical potential $\overline{\mu}$
along one-dimensional cuts in coupling space.
Figure \ref{fig:MF transitions}(a) shows
the dependence of $\overline{m}$,
$\overline{\chi}$,
and $\overline{\mu}$
on the electronic filling fraction $n^{\,}_{e}$
when $U=0.56$ and $V=0$ in units of $t^{\,}_{1}$.
All three are discontinuous functions of
$n^{\,}_{e}$ at two critical values of $n^{\,}_{e}$,
one below and another above $n^{\,}_{e,\mathrm{ms}}$,
for which the Stoner instability takes place.
Finite-size scaling is consistent with
a discontinuous dependence of $\overline{m}$ and $\overline{\chi}$
on $n^{\,}_{e}$ in the thermodynamic limit
upon entering the itinerant ferromagnetic phase.
The overlap of $\overline{\chi}$ and $\overline{\mu}$ 
is due to the fact that both are monotonically increasing functions 
of $n^{\,}_{e}-n^{\,}_{e,\mathrm{ms}}$
(except at their discontinuities) 
and their dependence can be approximated linearly
for small
$|n^{\,}_{e}-n^{\,}_{e,\mathrm{ms}}|\ll n^{\,}_{e,\mathrm{ms}}$.
Figure \ref{fig:MF transitions}(b) shows
the dependence of $\overline{m}$,
$\overline{\chi}$, and $\overline{\mu}$ 
on the nearest-neighbor interaction $V$
when $U=0.56$ in units of $t^{\,}_{1}$ and
$n^{\,}_{e}=n^{\,}_{e,\mathrm{ms}}-0.034$. 
Hereto, all three are expected from finite-size scaling to be
discontinuous functions of $V$
at the critical values of $V$ for which the Stoner instability takes place
in the thermodynamic limit.
The disappearance of the Stoner instability 
for large $V$ is due to the shift of the maximum of the
DOS to $n^{\,}_{e,\mathrm{min}}<n^{\,}_{e,\mathrm{ms}}$
as is implied by Fig.~%
\ref{fig:MF phase diag}(b). 
Increasing the values of $V$ holding $U$ and
$n^{\,}_{e}$ fixed with $n^{\,}_{e}<n^{\,}_{e,\mathrm{ms}}$
is effectively changing the DOS in a nonmonotonic way.
The DOS first increases, reaches a maximum, and then decreases
as a function of $V$. Correspondingly, if the given
values of $U$ and $n^{\,}_{e}$ are suitable in that
the maximum DOS is large enough for a Stoner instability to take place,
then increasing $V$ first triggers a Stoner instability followed
by a re-entrant phase transition
to the paramagnetic state when the DOS has decreased to a value 
too far from its maximum.
In contrasts to Fig.~%
\ref{fig:MF transitions}(a), 
$\overline{\mu}$ is an increasing function of $V$ while 
$\overline{\chi}$ is a decreasing function of $V$ (except at
their discontinuities).
The increase in chemical potential $\overline{\mu}$
can be understood as follows. The function $-V\,\overline{\chi}$
of $V$ is monotonically increasing.
Therefore, the renormalized hopping amplitude \eqref{eq:MF dispersion e}
is greater than its bare value, i.e., $\overline{t}^{\,}_{1,i}>t^{\,}_{1}$.
This results in an increase of both the bandwidths and
the gap between the $\tau=+$ and $\tau =-$ bands in such a way that a
greater $\overline{\mu}$ is required to keep
the filling fraction at $n^{\,}_{e}=n^{\,}_{e,\mathrm{ms}}-0.034$.

\begin{figure}[t]
\centering
\includegraphics[width=0.85\columnwidth]{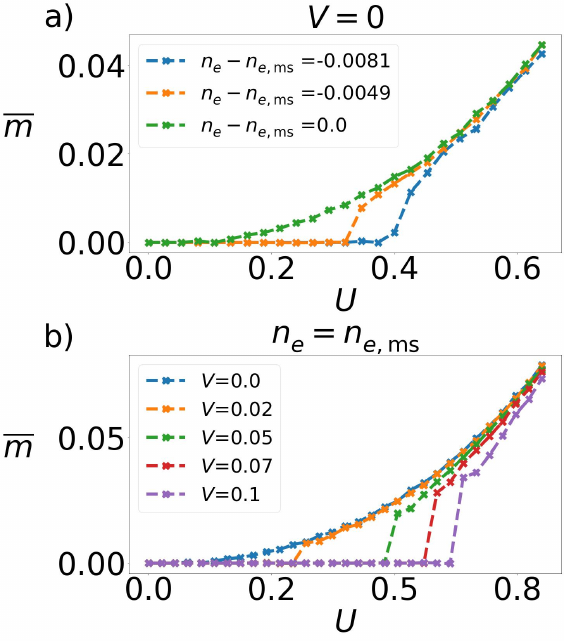}
\caption{
Dependency of the uniform magnetization $\overline{m}$
along one-dimensional cuts
in coupling space at zero temperature
as is explained in the text.
        }
\label{fig:MF Uc depencence}
\end{figure}

Figure \ref{fig:MF Uc depencence}(a) shows
the dependence of $\overline{m}$ on the on-site interaction $U$
when $V=0$ for different fixed values of $n^{\,}_{e}$.
The critical value $U^{\,}_{\mathrm{c}}$
for the onset of the Stoner instability
is minimal when
$n^{\,}_{e}=n^{\,}_{e,\mathrm{ms}}$.
It increases with the deviation
$|n^{\,}_{e}-n^{\,}_{e,\mathrm{ms}}|$.
Finite-size scaling is consistent
with $U^{\,}_{\mathrm{c}}$
vanishing when
$n^{\,}_{e}=n^{\,}_{e,\mathrm{ms}}$.
When $n^{\,}_{e}\neq n^{\,}_{e,\mathrm{ms}}$,
finite-size scaling is consistent with
$\overline{m}$ being a discontinuous
function of $U$ in the thermodynamic limit
upon entering the itinerant ferromagnetic phase
at $U^{\,}_{\mathrm{c}}>0$.
Figure~\ref{fig:MF Uc depencence}(b)
shows the dependence of $\overline{m}$ on the on-site interaction $U$
when
$n^{\,}_{e}=n^{\,}_{e,\mathrm{ms}}$
for different fixed values of $V$.
The critical value $U^{\,}_{\mathrm{c}}$
for the onset of the Stoner instability
is minimal when $V=0$.
It increases with increasing $V$. 
When $V>0$, finite-size scaling is consistent with 
$\overline{m}$
being a discontinuous function of $U$ in the thermodynamic limit
upon entering the itinerant ferromagnetic phase at $U^{\,}_{\mathrm{c}}>0$.

\subsection{Spinless Case}

In Sec.~%
\ref{subsec:MF Spinful case}, we showed 
for spinful electrons that the monkey-saddle singularity in 
the noninteracting limit is unstable against 
on-site Hubbard interaction at the mean-field level. 
We also
argued that the disappearance of the monkey-saddle singularity
when a repulsive nearest-neighbor interaction is present
is  due to the renormalization \eqref{eq:MF dispersion e}
of the bare hopping amplitude
$t^{\,}_{1}$.
A natural question that arises is the following.
Are there fine-tuned values of the couplings
$t^{\,}_{1}$, $t^{\,}_{2}$, and $M$ entering
the noninteracting dispersion \eqref{eq:Haldane two bands}
such that a monkey-saddle singularity is stabilized by a
repulsive nearest-neighbor interaction treated within mean-field
theory? 
Here, we will consider the case of spinless electrons
for which the on-site Hubbard term is not present and answer this 
question affirmatively.

To this end,
we consider the mean-field dispersion
\begin{subequations}
\label{eq:MF dispersion spinless}
\begin{equation}
\overline{\varepsilon}^{\,}_{\tau,\bm{k}}=
\tau\,\overline{\varepsilon}^{\,}_{\bm{k}},
\label{eq:MF dispersion spinless a}
\end{equation}
with
\begin{align}
&
\overline{\varepsilon}^{\,}_{\bm{k}}=
\sqrt{
M^{2}_{\bm{k}}
+
\left|
\Phi^{\,}_{\bm{k}}
\right|^{2}
},
\label{eq:MF dispersion spinless b}
\\
&
M^{\,}_{\bm{k}}=
M
+
2\,t^{\,}_{2}
\sum_{i=1}^{3}
\sin (\bm{k} \cdot \bm{b}_i),
\label{eq:MF dispersion spinless c}
\\
&
\Phi^{\,}_{\bm{k}}
=
\sum\limits_{i=1}^{3}
\overline{t}^{\,}_{1}\,
e^{+\mathrm{i}\bm{k}\cdot\bm{a}^{\,}_{i}},
\label{eq:MF dispersion spinless d}
\\
&
\overline{t}^{\,}_{1}=
t^{\,}_{1}
+
\delta
-
V\,\overline{\chi},
\label{eq: introduction delta in bar t1}
\end{align}
\end{subequations}
which is the mean-field dispersion \eqref{eq:MF dispersion}
where we set $U$, $\overline{m}$, and $\overline{\chi}^{\,}_{1}$
to be zero and removed the spin index $\sigma$.
Here,
$\delta$ is a tunable parameter that encodes the deviations from 
$t^{\,}_{1}$.
We retain the bare values of $t^{\,}_{2}$ and $M$
measured in units of $t^{\,}_{1}$
in the mean-field dispersion \eqref{eq:MF dispersion}.
Hence, when $\delta=V=0$, a monkey-saddle singularity is 
present in the dispersion at the filling fraction
$n^{\,}_{e,\mathrm{ms}}/2$. (Here, the division by $2$ 
is due to the removal of half of the bands for the spinless electrons.) 
This is not true anymore 
for $\delta\neq 0$ and $V=0$ since $\overline{t}^{\,}_{1}$
differs from $t^{\,}_{1}$ so that the monkey-saddle condition
$M=M^{\,}_{0}$ is not met anymore if we substitute
$t^{\,}_{1}$ with $t^{\,}_{1}+\delta$ in
$M^{\,}_{0}$ given by Eq.\
(\ref{eq:definition of M0}).
Conversely,
the mean-field dispersion \eqref{eq:MF dispersion spinless}
is identical to the 
noninteracting dispersion \eqref{eq:Haldane two bands}
but with the substitution $t^{\,}_{1}\to \overline{t}^{\,}_{1}$.
Because $\delta$ is only shifting the value of $\overline{t}^{\,}_{1}$
while we keep $M$ and $t^{\,}_{2}$ fixed, a monkey saddle 
singularity is guaranteed to 
exist in the spectrum only when $\overline{t}^{\,}_{1}=t^{\,}_{1}$
and at the filling fraction $n^{\,}_{e,\mathrm{ms}}/2$.
With these assumptions for the mean-field
dispersion, we must solve for
$\overline{\mu}(V)$, $\overline{\chi}(V)$, and $\overline{\delta}(V)$
the three coupled and non-linear mean-field equations
\begin{subequations}
\label{eq:modified MF eqs}
\begin{align}
&
n^{\,}_{e,\mathrm{ms}}
=
\frac{1}{|\Lambda|}
\sum_{\bm{k}, \tau}
f^{\,}_{\mathrm{FD}}
\left(
\overline{\varepsilon}^{\,}_{\tau,\bm{k}}
-
\overline{\mu}
\right),
\label{eq:modified MF eqs a}
\\
&
\overline{\chi}
=
\frac{1}{2|\Lambda|}
\sum_{\bm{k}, \tau}
\tau
\mathrm{Re}
\left\{
e^{\mathrm{i}\bm{k}\cdot\bm{a}^{\,}_{2}}\,
\frac{\Phi^{*}_{\bm{k}}}{2\,\overline{\varepsilon}^{\,}_{\bm{k}}}
\right\}
f^{\,}_{\mathrm{FD}}
\left(
\overline{\varepsilon}^{\,}_{\tau,\bm{k}}
-
\overline{\mu}
\right),
\label{eq:modified MF eqs b}
\\
&
\overline{t}^{\,}_{1,i}=t^{\,}_{1}
\iff
\overline{\delta}
=
V\,\,\overline{\chi},
\label{eq:modified MF eqs c}
\end{align}
\end{subequations}
as a function of the repulsive nearest-neighbor interaction strength $V$.
Solutions to Eq.~%
\eqref{eq:modified MF eqs} identify for which 
fine-tuned values $\overline{\delta}$ of the parameter $\delta$,
a monkey-saddle singularity is stabilized by a
repulsive nearest-neighbor interaction $V$ treated within a
mean-field approximation.
Notice that for any given values of the parameters $n^{\,}_{e}$, $V$,
and $\delta$,
Eqs.~%
\eqref{eq:modified MF eqs a} and~%
\eqref{eq:modified MF eqs b}
always have a solution. However, a monkey-saddle singularity
is present in the mean-field dispersion at the energy $\overline{\mu}$
only when $n^{\,}_{e}=n^{\,}_{e,\mathrm{ms}}/2$
and Eq.~%
\eqref{eq:modified MF eqs c}
is satisfied.

\begin{figure}[t]
\centering
\includegraphics[width=\columnwidth]{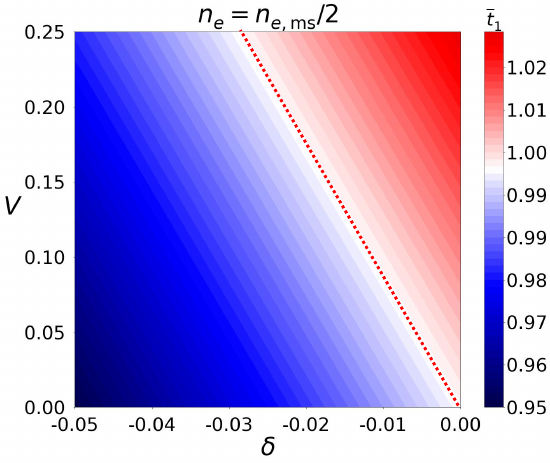}
\caption{
Renormalized hopping amplitude $\overline{t}^{\,}_{1}$
that is obtained by solving the self-consistent mean-field equations 
\eqref{eq:modified MF eqs} at zero temperature.
The red dashed line shows the points in the parameter
space for which the condition
\eqref{eq:modified MF eqs c}
is met. For any interaction strength $V$, there is a fine-tuned value
$\bar{\delta}$
of $\delta$
for which an interacting monkey-saddle singularity appears in the mean-field 
spectrum at the filling $n^{\,}_{e}=n^{\,}_{e,\mathrm{ms}}/2$.
}
\label{fig:MF_intMS}
\end{figure}

In Fig.~%
\ref{fig:MF_intMS}, we fix the filling fraction 
to $n^{\,}_{e}=n^{\,}_{e,\mathrm{ms}}/2$ and plot the 
renormalized hopping amplitude $\overline{t}^{\,}_{1}$ that is obtained 
by solving the mean-field equations \eqref{eq:modified MF eqs a}
and \eqref{eq:modified MF eqs b}
in parameter space of $\delta$ and $V$.
We find that, at the mean-field level and
for any given interaction strength $0\leq V\leq0.25$,
there exists a fine-tuned value $\bar{\delta}$ of $\delta$ for which 
an interacting monkey-saddle singularity appears at the filling
$n^{\,}_{e}=n^{\,}_{e,\mathrm{ms}}/2$.
Notice that for fixed $\overline{t}^{\,}_{1}$ at the filling
$n^{\,}_{e}=n^{\,}_{e,\mathrm{ms}}/2$, the solution
to Eq.~%
\eqref{eq:modified MF eqs b} fixes 
the value of $\overline{\chi}$. Equation
\eqref{eq: introduction delta in bar t1} then implies a linear relation 
between $V$ and $\delta$. In other words, constant $\overline{t}^{\,}_{1}$
contours in the $V-\delta$ plane must necessarily be linear
as is the case in Fig.~%
\ref{fig:MF_intMS}. 
We show the linear contour for which Eq.~%
\eqref{eq:modified MF eqs c}
is solved by the red dashed line in Fig.~%
\ref{fig:MF_intMS}.

\subsection{Conclusions}

To recapitulate, for the spinful electrons
mean-field theory predicts that the noninteracting
monkey-saddle singularity is unstable to both the
repulsive on-site Hubbard interaction and the
repulsive nearest-neighbor interactions for any
nonvanishing values of their coupling strengths.

In the former case, a Stoner
instability occurs for any nonvanishing $U$ at
$n^{\,}_{e}=n^{\,}_{e,\mathrm{ms}}$,
which destroys the monkey-saddle
singularity by
a rigid mean-field energy shift of the spin-up band
relative to that of the spin-down band.
Because of the itinerant ferromagnetic order,
spin-rotation symmetry is spontaneously broken.

In the latter case,
any finite coupling $V>0$ leads to a renormalization of the hopping
amplitude
$t^{\,}_{1}$ to
$\overline{t}^{\,}_{1,1}=
\overline{t}^{\,}_{1,2}=
\overline{t}^{\,}_{1,3}>
t^{\,}_{1}$.
This leads to a nonvanishing
$k^{2}$-correction to the monkey-saddle dispersion
\eqref{eq:Haldane two bands} that removes the higher-order
singularity through a Lifshitz transition of the Fermi surface.

We then showed that this 
removal of the monkey-saddle singularity  when $V>0$
can be compensated by the fine-tuning of
the bare hopping amplitude $t^{\,}_{1}$ such that 
an interacting monkey-saddle singularity appears in the mean-field dispersion.
For the spinless electrons, this fine-tuned interacting 
monkey-saddle singularity is stable as the on-site Hubbard interaction 
is inactive.

\section{Summary} 
\label{sec: Summary}

We addressed the question of whether it is possible to obtain single
odd higher-order singularities in the dispersion of an electronic
system. The motivation for this search is that when singularities
appear in pairs, interactions naturally lead to instabilities towards
ordered phases because of the scattering between each of the members
of the pair of singularities. In contrast, the types of instabilities
that can occur for isolated singularities are limited, and therefore
could potentially lead to non-Fermi-liquid behavior\,
\cite{monkey_saddle, supermetal}. While even singularities may occur
in systems where time-reversal symmetry is present, this symmetry
forbids odd singularities, such as a monkey saddle, to appear alone
inside the Brillouin zone. Here we showed explicit examples where odd
singularities may appear in isolation once time-reversal symmetry is
broken. The simplest example is perhaps the Haldane model, where we
find that varying a staggered chemical potential yields a single
monkey saddle singularity at one of the $\mathbf{K}$ points of the
hexagonal Brillouin zone, at an energy that sits within a gap 
with respect to momenta near the other (opposite) $\mathbf{K}$ point.

We then turned our attention
to the effects of interactions for an isolated odd
monkey-saddle singularity. Renormalization group flows inform us that
the interactions are relevant\,
\cite{monkey_saddle},
but do not identify the fate of the electronic state
when the chemical potential
is placed at the value where the Fermi surface changes
its topology. 
We carried out a mean-field calculation, including on-site and
nearest-neighbor interactions, that resolves the fate of the
monkey-saddle singularity.

For the case of spinful electrons, we 
obtained two phases as a result of the addition of
these interactions. One is a paramagnetic phase in which the
interactions lead to a deformation of the Fermi surface that avoids
the singularity. Basically, the system avoids the divergent DOS
through a renormalization of the 
nearest-neighbor hopping amplitude that redraws the shape of the
Fermi surface without breaking any lattice symmetry.
The other phase is an itinerant
ferromagnet, i.e., with Fermi surfaces
of different topology for the up- and down-spin species. 
This case is particularly interesting in that quantum oscillations of
magneto-resistance would reveal two different periods 
for Shubnikov--de Haas oscillations associated with the up
and down spins that differ by a factor close to 3.

In contrast to the spinful case, 
we have shown for spinless electrons that, 
in the presence of short-range repulsive interaction
that are treated at the mean-field level,
a monkey-saddle singularity can be stabilized by fine tuning
the hopping amplitudes.

As opposed to van Hove singularities, monkey-saddle singularities
do not generically appear. 
Instead, they require the fine tuning of at least one parameter 
in addition to the chemical potential in noninteracting 2D Hamiltonians. 
We have shown that, by fine tuning two parameters
in a spinless 2D Hamiltonian with nearest-neighbor interactions, 
one can obtain 
a monkey-saddle singularity. 
Recent experimental research efforts have been 
directed at increasing the number of 
continuously tunable parameters in 2D materials, 
most prominently in van der Waals materials. 
Such parameters include magnetic field, displacement field, and twist angles. 
It is thus opportune to look for monkey-saddle physics in these materials.

While these instabilities resolve the fate of the singularity in the
presence of interactions, there is a regime of temperatures for which
the quasiparticle lifetimes should display non-Fermi liquid behavior,
up to the low temperature scale for which the instabilities occur. In
all, both these intermediate regimes, as well as the interesting
signatures of the instabilities due to the multiple
Fermi-surface topologies and geometries that result
from interactions, make these systems rather rich,
and worthy of further investigations.

\section*{Acknowledgments}
\"OMA is supported by the Swiss National
Science Foundation (SNSF) under Grant No. 200021 184637.
AT is supported by the Swedish Research Council (VR) 
through grants number 2019-04736 and 2020-00214.
TN acknowledges support from the European Union’s Horizon 2020 research 
and innovation program (ERC-StG-Neupert-757867-PARATOP). 
CC acknowledges the support from the DOE Grant No. DE-FG02-06ER46316. 
AC acknowledges support the EPSRC Grant No. EP/T034351/1.
 
\bibliography{refs}
\bibliographystyle{apsrev4-2}
\end{document}